\documentclass[twocolumn]{aa}
\usepackage{mathtools}
\usepackage{amsmath}

\makeatletter
\@ifundefined{date}{}{\date{}}
\usepackage{txfonts}
\usepackage{epic}
\usepackage{eepic}
\usepackage{graphicx}
\usepackage{epstopdf}
\usepackage{xcolor}
\usepackage{todonotes}
\usepackage{todonotes}
\usepackage{subfig}
\setcitestyle{notesep={ }}

\def\ii{\textrm{i}}
\def\Cref{\ensuremath{\overline{C}^{\ \textrm{ref}}}}

\makeatother

\begin{document}

\title{Solar east-west flow correlations that persist for months at low latitudes are dominated by active region inflows}
\titlerunning{Solar east-west flow correlations are dominated by active region inflows}

\author{Chris~S.~Hanson \inst{1}, Thomas~L. Duvall Jr. \inst{2}, Aaron C. Birch\inst{2}, Laurent Gizon\inst{2,3,1} and Katepalli R. Sreenivasan\inst{4,1}}
\authorrunning{Hanson et. al.}

\institute{\inst{1} Center for Space Science, NYUAD Institute, New York University
Abu Dhabi, Abu Dhabi, UAE \\
   \inst{2} Max-Planck-Institut f\"ur Sonnensystemforschung, Justus-von-Liebig-Weg 3, 37077 G\"ottingen, Germany\\
   \inst{3} Institut f\"ur Astrophysik, Georg-August-Universit\"at G\"ottingen, Friedrich-Hund-Platz 1, 37077 G\"ottingen, Germany\\
   \inst{4} Department of Physics, Courant Institute of Mathematical Sciences, Tandon School of Engineering, New York University, New York, 10003\\
 \email{hanson@nyu.edu} }

\abstract{
{Giant-cell convection is believed to be an important component of solar dynamics. For example, it is expected to  play a crucial role in maintaining the Sun's differential rotation}. 
}
{We re-examine {early} reports of giant convective cells {detected using correlation analysis of Dopplergrams.}
{We extend this analysis} 
using 19 years of space and ground-based observations {of near-surface horizontal flows}. 

}
{
Flow maps are derived through local correlation tracking of granules and helioseismic ring-diagram analysis. {We compute temporal} auto-correlation functions of the east-west flows {at fixed latitude}. 
}{
{
Correlations in the east-west velocity can be clearly seen up to five rotation periods.
The signal consists of features with longitudinal wavenumbers up to $m=9$ at low latitudes. Comparison with magnetic images indicates that these flow features are associated with  magnetic activity.  
The signal is not seen above the noise level during solar minimum.}
}
{
{Our results show that the long-term correlations in east-west flows at low latitudes are dominantly due to inflows into active regions and not to giant convective cells.} 

}

\keywords{Sun: helioseismology - Sun: oscillations - Sun: activity - Sun: interior - convection - waves}

\maketitle


\section{Introduction}
The transport of heat from the solar interior to the surface is achieved through convective motions in the outer 30\% of the Sun. 
From observations, we currently know of two  well defined scales of convection; granulation and supergranulation. Granulation cells have lifetimes of $\sim$10~minutes, scales of 1-2~Mm ({{spherical harmonic degree}} $\ell\ge1000)$ and are well reproduced by numerical studies \citep[see review of ][]{nordlund_etal_2009}. Supergranulation is a larger scale of convection at $\sim$30~Mm $(\ell\sim120)$, with lifetimes of one to two days  and a depth structure that is debated \citep[see][for a review]{rincon_rieutord_2018}. The nature of convective motions at larger scales $(\ell<100)$ is not well understood, with a number of studies reporting different amplitudes \citep[see review of ][]{hanasoge_etal_2016} and different geometries \citep{beck_etal_1998,hathaway_etal_2013}. Here we reexamine the earliest reported detection of giant convective cells by \citet{beck_etal_1998}, who used a 16 month time series of Dopplergrams from space, by utilizing the subsequent two decades of observations.

{
In examining surface Doppler images from Mount Wilson observatory, \citet{labonte_etal_1981}, and later \citet{snodgrass_howard_1984}, concluded that there was no giant cell signal above 1-10~m/s per wavenumber. It was not until the availability of data from the Michelson Doppler Imager (MDI) on board the Solar and Heliospheric Observatory spacecraft that \citet{beck_etal_1998} detected long-lived large-scale velocity features. These features were found to have  an e-folding lifetime of 126~days and to be highly elongated in longitude (up to $44^\circ$), with an aspect ratio  $>$4 on the surface. \citet{beck_etal_1998} attributed this signal to the presence of giant convective cells. This result has been in contradiction with many numerical studies, which suggest that large-scale convection should form `banana-cells' that are elongated in latitude \citep[e.g.][]{miesch_etal_2008}. {\citet{ulrich_2001} disputed the convective nature of the observed EW flow correlations and proposed that they are Rossby or inertial waves.}}

Recent studies based on either the granulation tracking, supergranulation tracking, or local helioseismology have identified several components of near-surface large-scale motions:
(1) surface inflows into active regions \citep{gizon_etal_2001,gizon_2004, haber_etal_2004},
(2) vortical motions due to equatorial Rossby waves below $\sim 30^\circ$ latitude with longitudinal wavenumbers $m < 15$  \citep{loeptien_etal_2018,hanasoge_mandal_2019,liang_etal_2019,proxauf_etal_2020,hanson_etal_2020}, and (3) east-west velocity features with $m=1$ above $50^\circ$ latitude \citep{hathaway_etal_2013,bogart_etal_2015, hathaway_upton_2020}.

While the aforementioned studies focused on imaging distinct patterns of motion, a number of studies focused on constraining the amplitude of the flows. \citet{hanasoge_etal_2012} used local helioseismic techniques to measure the horizontal flow velocities at a depth of $0.96$R$_\odot$, finding {upper limits on the} amplitudes of the order 0.1 m~s$^{-1}$ per mode. This suppression of power at large scales compared to numerical simulations is puzzling \citep[see][for a commentary]{gizon_birch_2012} considering numerical studies had anticipated amplitudes two orders of magnitudes larger \citep{miesch_etal_2008}. To complicate the debate, \citet{greer_etal_2015} used a different local helioseismic technique and reported amplitudes larger than numerical studies  for large-scale features ($\ell<30$), but comparable amplitudes at smaller scales ($\ell>30$). For a careful assessment of the issues involved in these studies, see \citet{proxauf_thesis_2020}.

\begin{figure}
    \centering
    \includegraphics[width=\linewidth]{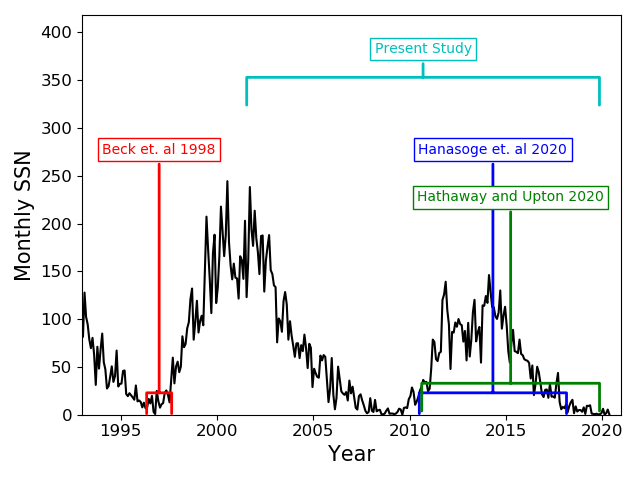}
    \caption{The monthly sunspot number (SSN) for cycle 23 and cycle 24. For reference, the observation period of the present study and references herein are highlighted.}
    \label{fig.monthlySSN}
\end{figure}

In this study we will address the observations of \citet{beck_etal_1998}. With an additional 19~years of observations (spanning nearly two solar cycles, see Fig.~\ref{fig.monthlySSN}), we seek to differentiate signals associated with magnetic activity, and those associated with non-magnetic processes (Rossby waves and possibly thermal convection). We will find that the signals identified by \citet{beck_etal_1998} are very likely due to the active region inflows, even near solar minimum.

\section{Data Analysis and results}

We use the horizontal flow maps derived from the observations of the ground-based Global Oscillation Network Group \citep[GONG,][]{Harvey_etal_1996} and the space-based Helioseismic and Magnetic Imager \citep[HMI,][]{schou_etal_2012a}. From the former, we utilize the 19~years of horizontal flows derived from the GONG++  ring-diagram analysis (RDA) pipeline from 2001 - 2020\footnote{{https://gong.nso.edu/}}. From the latter instrument, we use 8~years of RDA data from 2010-2018\footnote{{http://jsoc.stanford.edu/}}, and 6~years (2010-2016) of surface horizontal flow data derived from the local correlation tracking (LCT) of granules by \citet{loeptien_etal_2018}\footnote{{Data available upon request}}. For both RDA data sets we use the flows derived from $15^\circ$ tiles. To remove small-scale features from the LCT maps, we smooth them by convolving with a Gaussian of $\sigma=7.2^\circ$. 

In total we have three independent data sets. \citet{cobard_etal_2003} details the GONG++ RDA pipeline, \citet{bogart_etal_2011a,bogart_etal_2011b} details the HMI RDA pipeline and details of the LCT method are found in \citet{welsch_etal_2004} and \citet{fisher_welsch_2008}. For each of these data sets we have both of the horizontal flow components; in the direction of rotation $u_x$ and in the direction of the solar north pole $u_y$. We have corrected for center-to-limb effects through the method described by \citet{liang_etal_2019}. For clarity in terminology we refer to $u_x$ or $u_y$ as the measured flow within a single tile, and East-West (EW) or North-South (NS) as the flow signals averaged over longitude or temporal window.

\subsection{Coordinate system}
We examine the flows in the Stonyhurst coordinate system, whereby the frame is rotating at the Earth's orbital frequency. In this system, the latitude $\lambda$ and longitude $\phi$ are zero at the intersection of the equator and the central meridian on the visible disk. The latitude increases in the direction of the solar north pole, while the longitude increases in the direction of solar rotation. Accordingly, the longitudinal flows $u_\phi$ are positive if in the direction of rotation, and the latitudinal flows $u_\lambda$ are positive if in the direction of the north pole. The flow components are computed from small patches on the solar disk, assuming a Cartesian geometry ($x,y$). As such, within each small patch the approximation $u_x=u_\phi$ and $u_y=u_\lambda$ is taken.  {In this study we also analyze the horizontal divergence $\nabla_h\cdot\pmb{u}$ in spherical geometry, where $\nabla_h=(r\cos\lambda)^{-1}(\partial_\lambda\cos\lambda,\partial_\phi)$ and $\pmb{u}=(u_\lambda,u_\phi)$.}

\subsection{Auto-correlation functions}

Following \citet{beck_etal_1998}, we compute the auto-correlation functions of the flow maps. For each time segment of duration $N_t=720$~days, we consider the centered and normalized velocity components 
\begin{equation}
\bar{u}_i(\lambda,\phi,t) = \frac{{u}_i(\lambda,\phi,t) - \mu_i(\lambda,\phi)}{\sigma_i(\lambda,\phi)}
\end{equation}
where $\mu_i(\lambda,\phi)$ is mean value of $u_i$ over the 720~day window and $\sigma_i(\lambda,\phi)$ is standard deviation.
The longitude-averaged auto-correlation of the time-series $\bar{u}_x$ at fixed latitude at time lag $\tau$
is
\begin{equation}\label{eq.autocorr}
    \mathcal{A}_{\rm EW}(\lambda, \tau) = \frac{1}{N_t N_\phi}\sum_\phi \sum\limits_{t}  \bar{u}_x(\lambda, \phi, t)\bar{u}_x(\lambda, \phi, t+\tau). 
\end{equation}
The auto-correlation of $\bar{u}_y$ is
\begin{equation}\label{eq.autocorr_uy}
    \mathcal{A}_{\rm NS}(\lambda, \tau) = \frac{1}{N_t N_\phi}\sum_\phi \sum\limits_{t}  \bar{u}_y(\lambda, \phi, t)\bar{u}_y(\lambda, \phi, t+\tau).
\end{equation}

Long-lived features will appear as peaks in the auto correlation at integer values of the local rotation period. The auto correlations are computed using a temporal window of $\sim$720 days. Due to the different cadence of each of the data sets, this results in $N_t=633$ for each window in the RDA data and $N_t=720$ for the LCT data. For each consecutive sample the window is shifted by $N_t/2$.  In order to improve the signal-to-noise ratio, we then take the mean across longitude and temporal samples. When comparing LCT and RDA, we use the RDA auto correlation for flows at a depth of $\sim2$~Mm which are the shallowest {depth from the GONG++ full disk flow map pipeline}\footnote{{The depth grid is not consistent between GONG++ tiles. As such, the pipeline performs interpolation to obtain full disk flow maps at depths of 2, 4.5, 7 and 12.5~Mm. Flows at other depths require the user to interpolate into a chosen grid.}}. {Comparison between HMI RDA and LCT is possible at the surface, though \citet{proxauf_etal_2020} notes that HMI RDA surface measurements are unreliable. HMI RDA flow inversions that target the surface appear to have excess power at low $m$, relative to inversions 696~km deeper, and this results in large scale artifacts in the surface flow maps (Bastian Proxauf, private communication).}

\begin{figure}[!htb]
    \centering
    \subfloat[][]{\includegraphics[width=\linewidth]{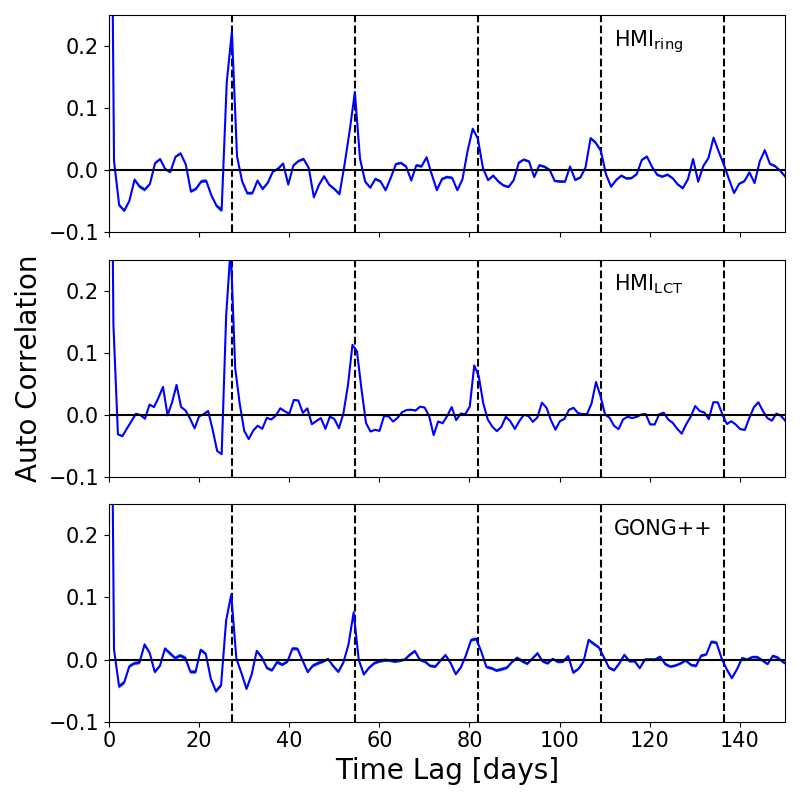}}\\
    \subfloat[][]{\includegraphics[width=\linewidth]{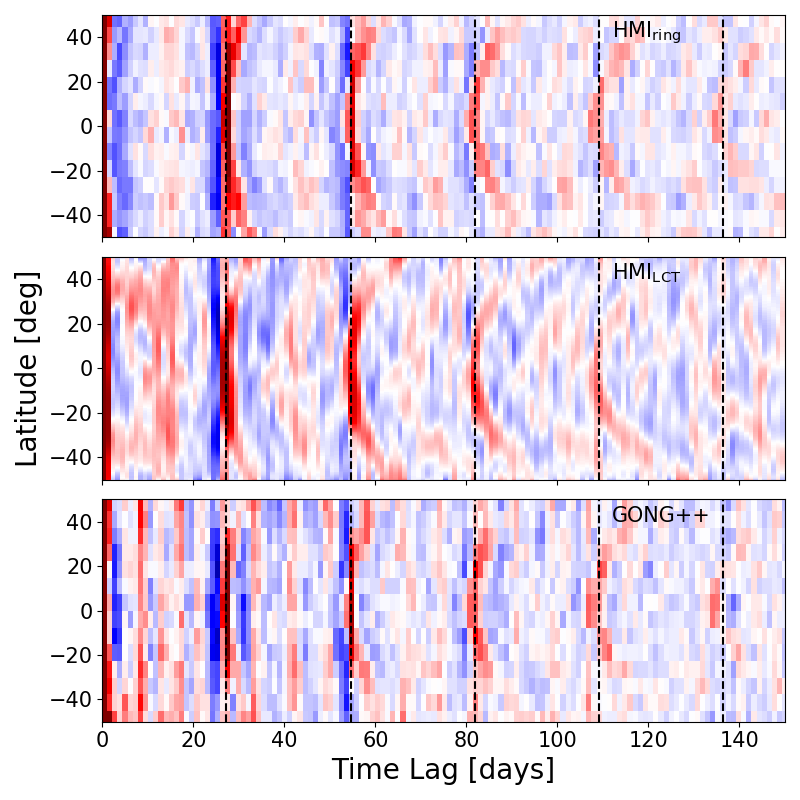}}\\
    \caption{Top: The temporal auto-correlation function of the EW flows at the equator for HMI RDA, LCT and GONG++ RDA. A signal is seen to persist for at least five solar rotations. Statistical error bars are comparable to line thickness, hence are not shown. Bottom: A 2D map of the EW auto-correlation maps, as a function of latitude and time lag. The deflection with latitude is qualitatively consistent with differential rotation. The dashed vertical lines in each image indicate multiples of the equatorial rotation period.}
    \label{fig.EW_Acorr_sat_comp}
\end{figure}

Figure~\ref{fig.EW_Acorr_sat_comp} shows the auto-correlations of the EW flows as a function of time lag and latitude. 
These results are consistent with \citet{beck_etal_1998}, showing a long-lived EW flow signal that persists for  at least five solar rotations and can be seen at latitudes up to $45^\circ$. Due to the {latitudinal} differential rotation, the peaks in $\mathcal{A}$ are at greater time lags at higher latitudes. In each data set, the peak in $\mathcal{A}$ is preceded and followed in time by negative correlation {sidelobes}. 

Figure~\ref{fig.NS_Acorr_sat_comp} shows similar results for the NS flow signal. However, in order to obtain such clear results the solar Rossby waves \citep{loeptien_etal_2018} were filtered out (see Appendix~\ref{sec.rossbyfilt}). While similar to the EW signal, the NS signal seems to become consistent with noise within four rotations.

{}

\subsection{Spectral Analysis}
Here, we examine the spectral components of the EW auto correlation in a similar manner to \citet{ulrich_2001}. For this section we present the results for HMI RDA. The results of the other data sets are similar.

We perform a temporal Fourier transform to the EW auto-correlation function (Eq.~\ref{eq.autocorr}),
\begin{equation}\label{eq.powerspectrum1}
    \hat{\mathcal{A}}_{\rm EW}(\lambda,\omega) =  \sum_\tau  \mathcal{A}_{\rm EW}(\lambda,\tau)\textrm{e}^{\ii \omega \tau} ,
\end{equation}
where the sum is taken over all time lags $\tau$, and $\omega$ is the angular frequency. 
  In the {Stoneyhurst} frame a flow pattern proportional to $e^{i m\phi}$ which is rotating at the rotation rate $\Omega_\lambda$  will have a frequency  $m\Omega_{\lambda}$, where $\Omega_{\lambda}$ is the local rotation rate at latitude $\lambda$. We have corrected the data for the Earth's orbital frequency 31.7~nHz.

\begin{figure}
    \centering
    \includegraphics[width=\linewidth]{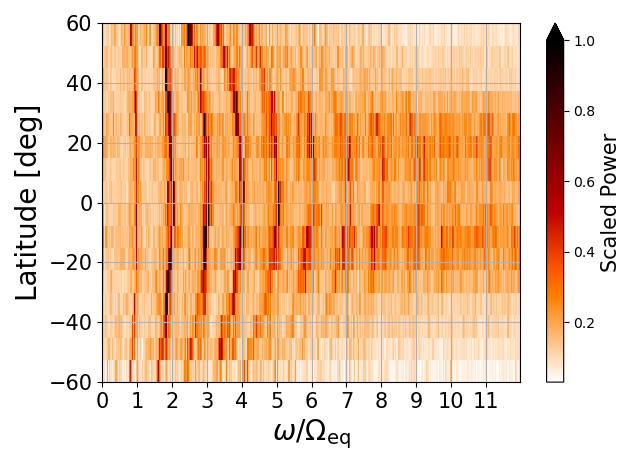}
    \caption{
   Plot of  $|\hat{\mathcal{A}}_{\rm EW}(\lambda,\omega)|^2$ for HMI RDA as a function of latitude and frequency. The frequency is scaled by the rotational frequency at the equator $\Omega_{\rm eq}$ ($2\pi$/27.275 days).
    At the equator, power can be seen near integer values of $\omega/\Omega_{\rm eq}$, up to values of about 9.  At latitudes above 30~deg, these higher-order modes are not visible. 
    }
    \label{fig.EW_Acorr_PowerSpectrum}
\end{figure}

Figure~\ref{fig.EW_Acorr_PowerSpectrum} shows $|\hat{\mathcal{A}}_{\rm EW}|^2$ 
as a function of frequency  and latitude, at a depth of 2~Mm.
These results show that $\mathcal{A}_{\rm EW}$ has components up to $m=9$, after which the signal is consistent with noise. At higher latitudes ($\lambda>40^\circ$) modes of order $m>6$ are not seen. The mode $m=2$ at the equator has the greatest amplitude. In the frequency domain, the deflections with latitude seen in Fig.~\ref{fig.EW_Acorr_sat_comp} are towards lower frequencies and are qualitatively consistent with differential rotation. 

\subsection{Mode parameters of the EW signal}

In this section we characterize the modes that contribute to the long-lived EW signal. We compute the power spectrum $P(\lambda,\omega)$  of the EW flow maps for the entire HMI observation time

\begin{equation}\label{eq.powerspectrum}
    P(\lambda, \omega) =   \left|\sum_\phi \sum_t  u_x(\lambda,\phi, t)\textrm{e}^{\ii \omega t} \right|^2,
\end{equation}
where the sum over $\phi$ is over the {range $-45^\circ\le \phi\le45^\circ$} for each snapshot.
The resulting spectra are similar to Fig.~\ref{fig.EW_Acorr_PowerSpectrum}, but have an exponential probability distribution function. 
{The azimuthal modes $m$ at each latitude $\lambda$ are fit assuming a Lorentzian line profile,
\begin{equation}\label{eq.lorenzFit}
    \mathcal{L}_{\lambda}(\omega,\pmb{\Lambda}_m) = 
    \frac{A_m}{1+(\omega - \omega_m)^2/(\Gamma_m/2)^2} + B_m,
\end{equation}{}
where $A_m$ is the mode amplitude, $\omega_m$ is the mode frequency, $\Gamma_m$ is the full width at half maximum, $B_m$ is the background noise of the spectrum and $\pmb{\Lambda}_m = \{A_m,\omega_m,\Gamma_m,B_m\}$. We fit only up to $m=6$ due to poor a signal-to-noise ratio for higher modes. Details of the fitting can be found in appendix~\ref{sec.modeFits}. We have opted to fit each individual mode $m$ using a frequency window of width $\Delta\omega/\Omega_{\rm \lambda}=1$.}

\begin{figure}
    \centering
    \includegraphics[width=\linewidth]{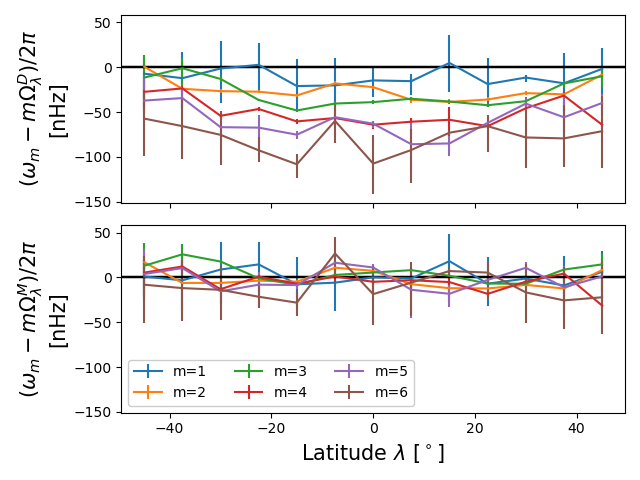}
    \caption{The difference between the rotational frequencies of the long-lived EW signals in HMI RDA and the synodic frequencies of Doppler features ($\Omega^D_\lambda$, top) or magnetic features ($\Omega^M_\lambda$, bottom). The Doppler and magnetic rotational frequencies are summarized in \citet{snodgrass_ulrich_1990}. Error bars are one sigma. 
    }
    \label{fig.powerSpectrum_fit_freq}
\end{figure}

\begin{figure}
\centering
    \subfloat[]{\includegraphics[width=\linewidth]{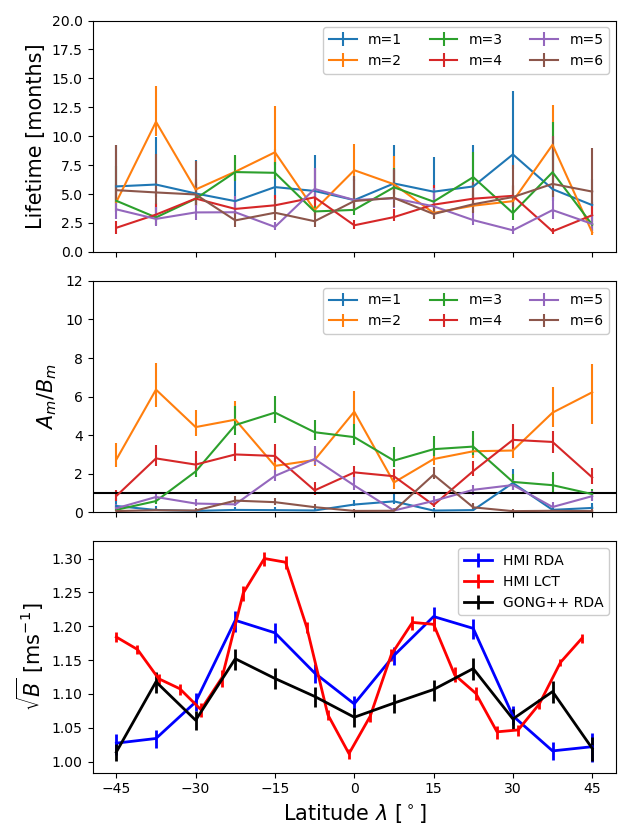}}\\
    \caption{Top: Lifetime of the modes seen in the HMI RDA EW power spectrum. Middle: the signal-to-noise ratio ($A_m/B_m$). Bottom: The background power $B=\sum_{m=1}^6B_m/5$ in the EW spectrum for all three data sets. Error bars are one sigma. 
    }
    \label{fig.powerSpectrum_fit}
\end{figure}

Having fit the power spectrum of the EW flows, we compare the rotational frequencies to those derived from Doppler and magnetic features \citep{snodgrass_ulrich_1990}. \citet{snodgrass_ulrich_1990} identified magnetic features from rebinned Mt. Wilson magnetograms, with each bin covering approximately $6^\circ$ in latitude. This coarse grid smooths out small-scale activity, and would be dominated by strong magnetic features. As such the derived rotation rates are consistent with sunspot rotation rates below $40^\circ$ latitude \citep[see review of][]{beck_2000}. 
Figure~\ref{fig.powerSpectrum_fit_freq} compares these rotation rates to the fitted rotation frequencies of the HMI RDA EW power spectrum. These results show a mismatch between the frequencies of the EW signal and Doppler features (supergranules). However, when compared to the magnetic tracers, the EW frequencies are similar at all latitudes. This indicates that the EW flow rotation rate is consistent with the rotation rate of magnetic regions. These results are also in agreement with HMI LCT and GONG++ RDA data sets.

Figure~\ref{fig.powerSpectrum_fit} shows the other fit parameters of the HMI RDA EW spectrum. {The measured lifetimes ($2/\Gamma_m$) show that within error estimates the modes at all latitudes have a lifetime of 4-5 months. This is consistent with \citet{beck_etal_1998}.} In the HMI RDA flow maps the signal-to-noise ratio SNR ($A_m/B_m$) is near or above parity (up to 6) for modes from $m=2$ up to $m=5$.
Finally, we compare the mean background power $(\sum_{m=1}^6 B_m/5)$ in the three data sets and find values of $\sim$1-1.2~m/s, with the greatest power at the active latitudes. This gives a limit on the sensitivity of detecting giant flow cells.

\subsection{Relation to Activity}
The results in the previous sections confirmed the presence of long-lived signals in the near-surface horizontal flows in three independent data sets. However, these results do not confirm the convective or magnetic nature of such flows. Rossby waves rotate retrograde at fractional values of $\omega/\Omega_{\rm eq}$ (in Stoneyhurst frame), and hence can be filtered from the data (see Appendix~\ref{sec.rossbyfilt}). 
At such large scales and long lifetimes, there remains {a few potential causes for the origin} of these signals; giant-cell convection, active-region inflows {or a side effect of magnetic activity}. 
With the long observation window of GONG++, we will exploit the data of two solar cycles to shed light on the nature of these flows.

To investigate this, we first perform further processing on the flow data sets and 
filter out modes greater than ${\omega}/\Omega_\lambda=8$ and less than ${\omega}/\Omega_\lambda=1$. This is done with a filter that has a flat top from ${\omega}/\Omega_\lambda=1$ to ${\omega}/\Omega_\lambda=8$, and is tapered to zero by a cosine bell that goes from 1 to zero in {a distance} $\Delta{\omega}/\Omega_\lambda=1$. This filter is applied to both negative and positive frequencies and the flow maps are then reconstructed into the time domain. Applying such a filter to the data introduces some smoothing and removes frequencies less than the equatorial rotation frequency.

For comparison with magnetic activity, we track MDI and HMI magnetograms, which overlap with the GONG++ data presented here. We utilize the tracking routine {mTrack}\footnote{http://hmi.stanford.edu/rings/modules/mtrack.html} which is the same routine used to track Doppler images for the HMI RDA pipeline. The magnetograms are sampled at a cadence of 96~minutes and tracked for 1/24\textsuperscript{th} of a Carrington Rotation (CR), which results in 17 magnetograms {per data cube}.  Each magnetogram is projected onto a cylindrical equidistant map, tracked at the Snodgrass magnetic tracer rotation rate \citep{snodgrass_ulrich_1990},  and sub sampled to a resolution of $0.6^\circ$ per pixel.

The flow and magnetic maps are collected into data cubes spanning one~year in time. We then identify {converging} flow features  with amplitudes $-\nabla_h\cdot\pmb{u} >  3\times10^{-8}$~s$^{-1}$, considering these signals to be above the background noise in the GONG++ divergence data. Here we focus on the convergent features, since the analysis of the resulting maps is intuitive when considering magnetic activity. We have performed the same analysis on EW and divergent flows, finding similar results. 
At fixed time, among the converging flow features that are within $30^\circ$ of each other we select the largest amplitude feature, to avoid counting a large converging flow features multiple times. 
Additionally, features outside of {40}$^\circ$ from disk center are rejected {to avoid the disk-edge}. On average we identify {850} features within each one~year window. For each identified feature, the flow and magnetic maps are shifted such that the feature is positioned at image center. A mean of the images is then computed. In order to determine the evolution of these flows, this procedure is repeated at integer (up to 3) CRs before and after the initial identification. Differential rotation $\Omega_\lambda^M$ is accounted for in this averaging routine.   

\begin{figure*}
    \centering
    \includegraphics[width=\linewidth]{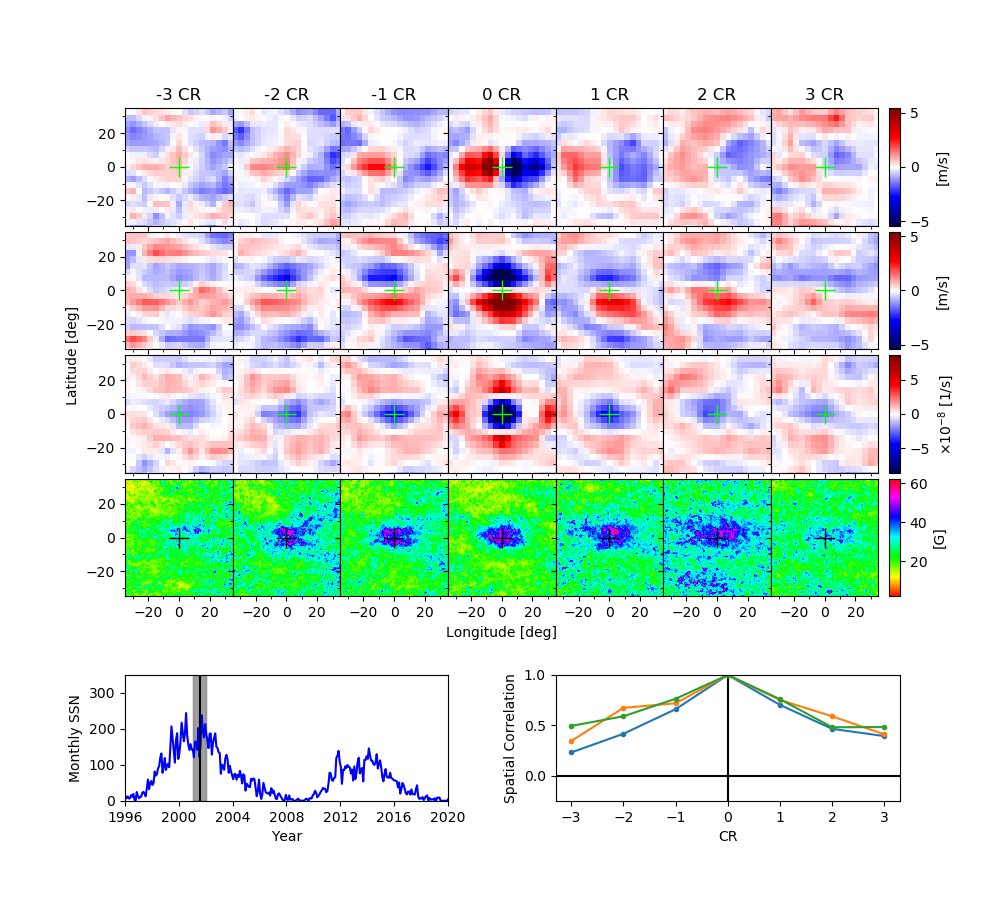}
    \caption{Top panels: Average flows around convergent features  during  solar maximum of cycle 23. Each of the first three rows is a different component of the flow, from top to bottom: EW flows, NS flows and divergence maps. The fourth row is for the average unsigned magnetic field.
    Each column is the mean map at an integer rotation before and after the initial identification (0 CR, middle column). 
    Bottom left panel: The monthly sunspot number as a function of year (blue), with the center of the 1 year observation window overplotted (vertical black line) and window size shown (shaded grey region). Bottom right panel: spatial correlation coefficient between the CR$=0$ flow maps and the flow maps at time lag CR (EW: blue, NS: orange, $\nabla_h\cdot\pmb{u}$: green).}
    \label{fig.DIV_stackedmaps_solarMax}
\end{figure*}

\begin{figure*}
    \centering
    \includegraphics[width=\linewidth]{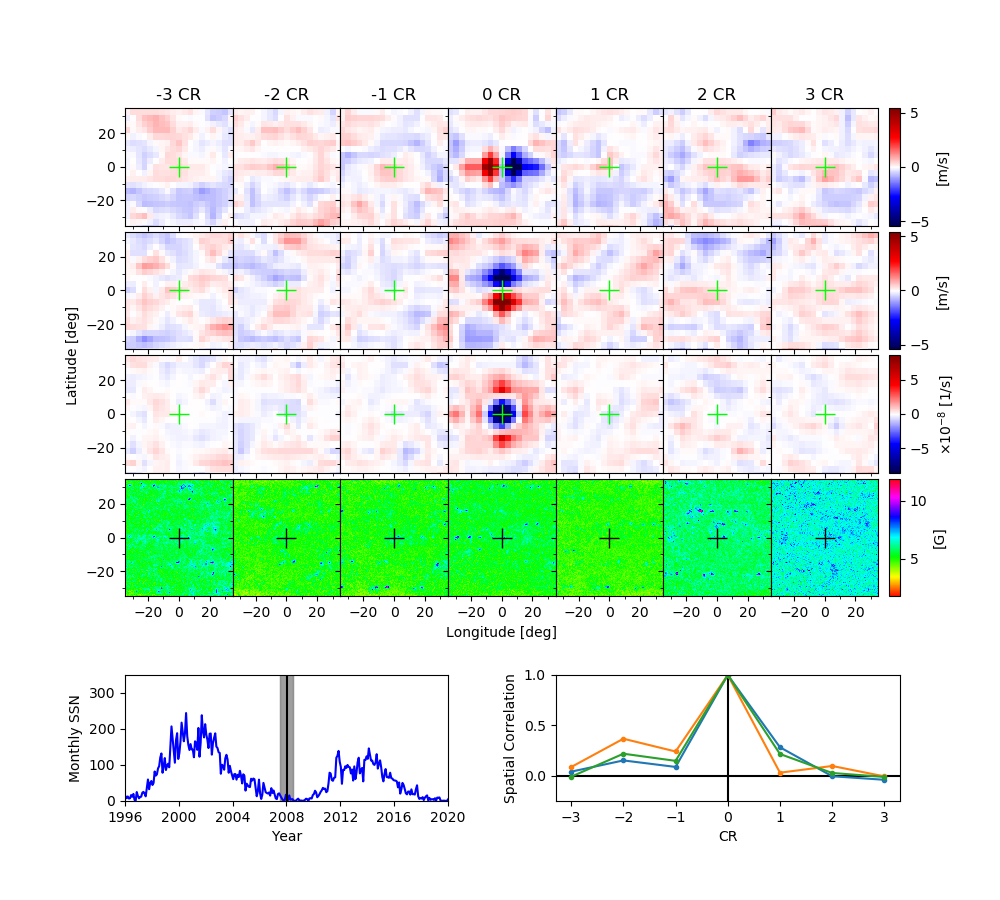}
    \caption{Same as Fig.~\ref{fig.DIV_stackedmaps_solarMax}, except at the solar minimum between cycle 23 and 24.}
    \label{fig.DIV_stackedmaps_solarMin}
\end{figure*}

Figures~\ref{fig.DIV_stackedmaps_solarMax} and \ref{fig.DIV_stackedmaps_solarMin} show these stacked maps for the solar maximum of cycle~23 and the minimum between cycles~23 and~24, respectively. For visualization we have linearly interpolated the RDA flow maps to improve resolution by factor of four. In both figures, large-scale flows can be observed at the initial identification time (CR=0). During the maximum, the EW and NS flows extend $30^\circ$ (from center) in the longitude direction and $15^\circ$ in latitude. These flows patterns are seen at least 3~CR before and after initial identification. The computed spatial correlation coefficients show that at $+$3~CR the flows have a correlation of $\sim0.45$ with the initial identified features. The corresponding magnetic maps show that these flows are related to activity. {We have not included the vorticity here, as the Rossby waves and Rossby filters make it difficult to obtain a clean signal at higher latitudes (see Appendix~\ref{sec.rossbyfilt} for discussion).}

During the minimum, the identified flows extend approximately $15^\circ$ from the center. The associated magnetic maps show very weak magnetic fields. Unlike the maximum, the identified flows in the minimum do not persist beyond a local rotation period. This is further evident in the correlation coefficients, dropping to near zero by $\pm1$~CR. The maps of the flows at integer rotations share no visual correlations with the identified feature.

\begin{figure}
    \centering
    \includegraphics[width=\linewidth]{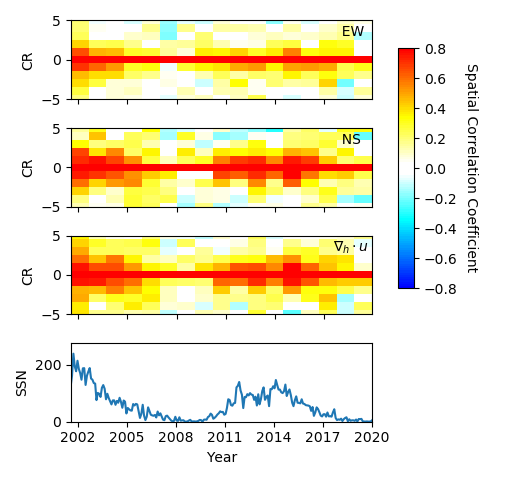}
    \caption{{The spatial correlation coefficient between flow maps at CR=0 and those at time lag CR, as a function of time and CR. The coefficients for the EW, NS and divergence are shown in the top three panels, with the monthly sunspot number shown for reference in the bottom panel. The large scale flows persist for a number of rotations during periods of high solar activity.}}
    \label{fig.spatialCorr}
\end{figure}

{Figure~\ref{fig.spatialCorr} shows the spatial correlation of the flow maps as a function of CR and time. These results further demonstrate that during solar maximum, these large scale flows persist for a number of rotations. As the solar cycle begins its descending phase, the flows persist for shorter periods of time. At the solar minimums, 2008 and 2019, the flows persist for no more than a rotation.  }

\section{Discussion and conclusions}

We have investigated the long-lived large-scale EW flow signal reported by \citet{beck_etal_1998}. Using nearly twenty years of flow maps derived from observations by both GONG and HMI, we confirm the presence of this signal in the near-surface horizontal flows. However, with careful comparisons with magnetic field maps, we find that these signals are consistent with magnetic active region inflows. Specifically, the rotation rate of these signals is consistent with the measured rotation rate of magnetic features \citep{snodgrass_ulrich_1990}. Furthermore, we find there are no long-lived features, above the noise level, observed to persist beyond a rotation during the solar minimum.

 \citet{beck_etal_1998} observed giant long-lived velocity features and attributed these to {convection}. This conclusion was based on the assumption that their data set, consisting of MDI Dopplergrams from 1996 to 1997, would not be significantly contaminated by magnetic activity during a minimum. While active region inflows were not discovered at this time, the authors were still aware that magnetic activity would affect the results. Nevertheless, during this observation window, cycle 23 was ascending and the monthly sunspot number was growing from 10 to 45 (see Fig.~\ref{fig.monthlySSN}). We investigated a similar period in cycle 24 (2009, 1 year after Fig.~\ref{fig.DIV_stackedmaps_solarMin}) and found  inflows into active regions  persisting for a number of months. These inflows are at a  weaker level than in Fig.~\ref{fig.DIV_stackedmaps_solarMax} and after 3 rotations have a spatial correlation coefficient of 0.25. The EW component of these inflows is what is observed in the correlation functions of \citet{beck_etal_1998}. We note that there is a very small window (1 year centered at the minimum) in which inflows do not dominate the low-latitude flow field. {Interestingly, the studies of \citet{labonte_etal_1981} and \citet{snodgrass_howard_1984} did not report on these long-time correlations, despite the data covering cycles 20 and 21 and the supposed sensitivity of $\sim  1$ m/s per wavenumber.}
 
 The reported observations of \citet{beck_etal_1998} are consistent with ours, with the exception that they incorrectly identified the large velocity cells as convective in origin. Specifically, the EW correlations have an e-folding lifetime of approximately 4~months, and the extent of the flows is greater in the longitude direction.
 \citet{beck_etal_1998} reported EW flows elongated in longitude with an aspect ratio greater than four.
 We also find EW flows elongated in longitudes,
 however with an aspect ratio around two (Fig.~\ref{fig.DIV_stackedmaps_solarMax}). We attribute this difference to the averaging scheme. We were careful to avoid selecting large  converging flow features multiple times to avoid smearing, thus our aspect ratio is smaller. The elongation of the spatial correlation of EW flows in the longitudinal direction is not surprising given the geometry of active regions. 

{In an argument against the convective nature of these cells, \citet{ulrich_2001} suggested that the long-lived signals in the EW flows might be due to Rossby or inertial waves. 
The solar equatorial Rossby waves, discovered and characterized subsequently,  are not seen in the EW flow maps at low latitudes and thus cannot cause the signals originally  observed  by  \citet{beck_etal_1998} and \citet{ulrich_2001}. We find that the active region inflows generate a flow pattern with azimuthal modes up to $m=9$, which is consistent with the spectral analysis of \citet{ulrich_2001}. 
\citet{ulrich_2001} found that while the rotational frequencies of these modes were consistent with magnetic features, a couple of modes were slower. 
However, in our study, we find that at each latitude all  modes have frequencies consistent with the rotational frequency  of magnetic features, and are absent during the solar minimum, thus indicating a magnetic association for these signals.
Aware of the potential affect in Doppler images, \citet{ulrich_2001}  defined quiet Sun as latitudes without magnetic activity. When studying solar dynamics, the definition of quiet Sun should be extended to exclude the latitudes that contain the inflows around active regions  \citep{gizon_etal_2001}. We've shown that the inflows extend $15^\circ$ in latitude from the center of active regions.}

\citeauthor{proxauf_thesis_2020} (\citeyear{proxauf_thesis_2020}, his figure 4.2) reported that during periods of high magnetic activity, the east-west flow amplitude at the surface increases by $\sim3$~m\ s$^{-1}$ for $\ell<10$. He attributed this change in the power spectrum to inflows into active regions, which is consistent with the present study.
 
 Meanwhile \citet{hanasoge_etal_2020} examined toroidal flow power as a function of depth, latitude and wave number, and compared observations to simulations. These authors showed that the observed toroidal flow is higher in low latitude regions than at high latitudes, which is in contrast to simulations. We note that the analysis of \citet{hanasoge_etal_2020} was performed from 2010 to 2018, covering cycle 24 (see Fig.~\ref{fig.monthlySSN}). Our analysis shows that large scale active region inflows will contribute to the toroidal (EW) component of the flow field during this period.

In future work, we intend to extend this study beyond $50^\circ$ latitudes, in order to investigate the dependency of the high-latitude flows  \citep{hathaway_etal_2013,bogart_etal_2015,hathaway_upton_2020}
with solar activity.

\appendix
\section{North south auto-correlation maps}\label{sec.rossbyfilt}

\begin{figure}
    \centering
    \includegraphics[width=\linewidth]{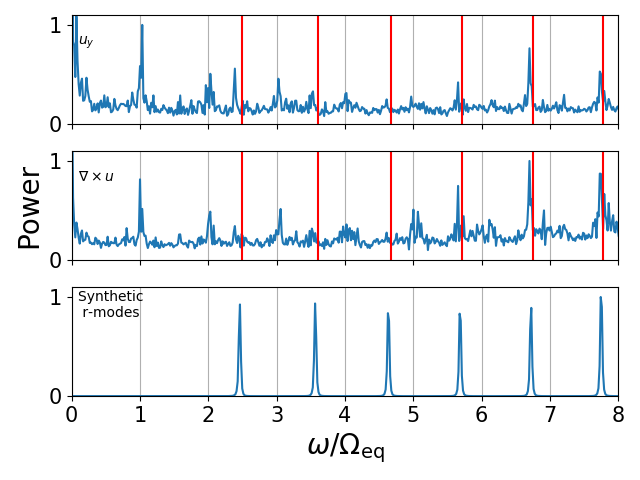}
    \caption{Power spectra of the  GONG++ flow components and synthetic Rossby waves in the Stoneyhurst coordinate system at the equator. From top to bottom panel are the $u_y$ flows, radial vorticity and $u_y$ component of synthetic Rossby waves \citep[e.g.][]{hanson_etal_2020}. Modes rotating at frequencies of multiples of $\Omega_{\rm eq}$ are indicated by the vertical grey lines. The vertical red lines show the rotational frequencies of the sectoral Rossby waves. We have considered only the $m>2$ sectoral Rossby waves seen in the Sun. }
    \label{fig.synRossby}
\end{figure}

It is well established that Rossby modes are present within the HMI LCT \citep{loeptien_etal_2018}, HMI RDA \citep{proxauf_etal_2020} and GONG++ RDA \citep{hanson_etal_2020} data. {
In the Stoneyhurst frame, Rossby waves will have a frequency of $\omega = \Omega_\textrm{eq}(m-2/(m+1))$.} Without accounting for them, the Rossby modes will contribute to the $u_y$ and radial vorticity maps. Figure~\ref{fig.synRossby} shows how the Rossby waves and the long-lived modes of this paper are both present in the NS and vorticity spectra. In order to remove the Rossby components, we remap the flows into a longitude-latitude frame that rotates at the equatorial rotation rate ($\Omega_\textrm{eq}/2\pi=453.1$~nHz). The flow maps within the equatorial rotation frame $\pmb{u}^{\textrm{eq}} = \{u_x^{\textrm{eq}},u_y^{\textrm{eq}}\}$ are then Fourier transformed in longitude and time,
\begin{equation}
    \pmb{u}^{\textrm{eq},m}(r,\theta,\omega) = \int\limits_0^T\int\limits_0^{2\pi}\pmb{u}^\textrm{eq}(r,\theta,\phi,t)\exp^{-\ii m\phi+\ii\omega t}\textrm{d}t\textrm{d}\phi,
\end{equation}
where $T$ is the total observation time of the respective data set. A cosine bell filter is then applied to each $m$, centered on the theoretical dispersion $\omega_r =-2\Omega_\textrm{eq}/(m+1)$ with a tapering length of $35$~nHz. The width $\mathcal{W}(m)$ of the filter follows,
\begin{equation}
    \mathcal{W}(m)=
    \begin{cases}
    75~\textrm{nHz} & m \leq 10 \\
    50~\textrm{nHz}& m > 10 \\
    \end{cases}
\end{equation}
where the width narrows for $m>10$ in order for the filter not to cross the zero frequency. We perform this filtering on $\pm3$ aliases from the central ridge (which arise from the window function), as well as in the $-m$ half-space.

\begin{figure}[!htb]
    \centering
    \subfloat[][]{\includegraphics[width=\linewidth]{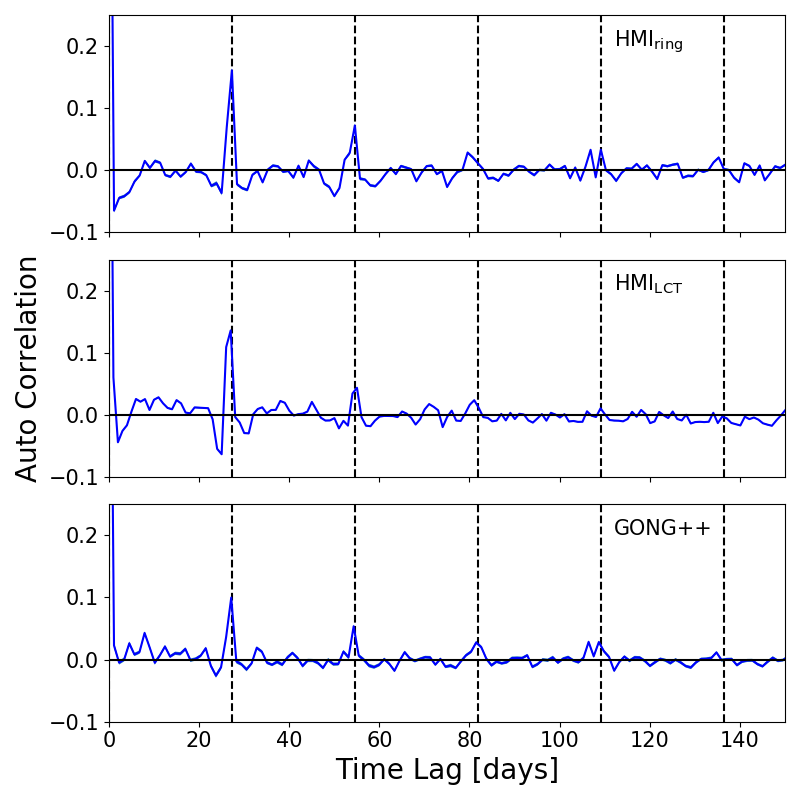}}\\
    \subfloat[][]{\includegraphics[width=\linewidth]{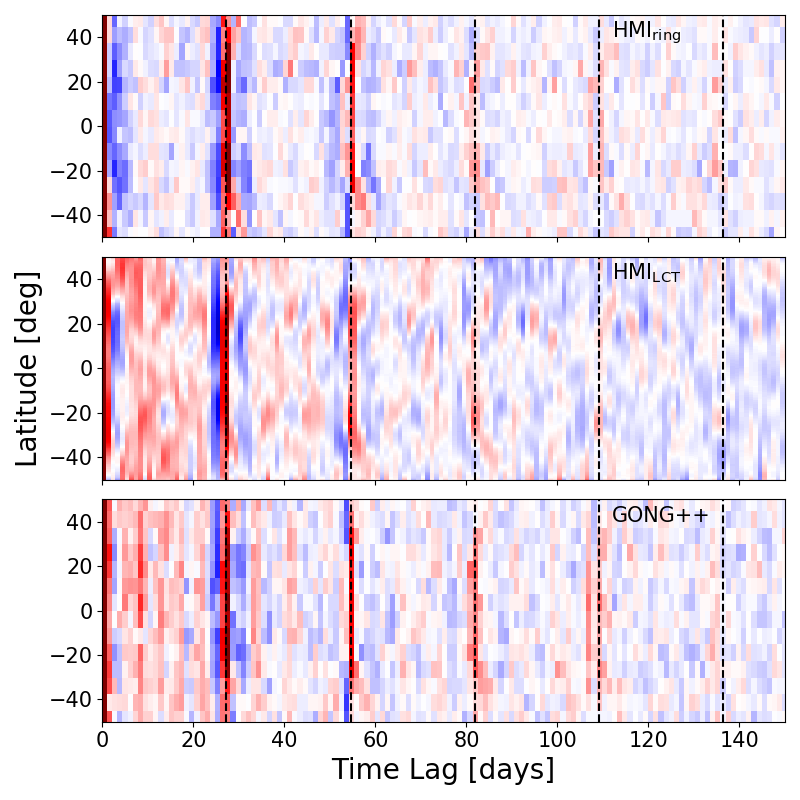}}\\
    \caption{Top: The temporal auto-correlation function of $u_y$ for HMI RDA, LCT and GONG++ RDA. A signal can be seen to persist for at least four solar rotations. Statistical error bars are comparable to line thickness, hence are not shown. Bottom: A 2D map of the $u_y$ auto-correlation maps, as a function of latitude and time lag. The deflection with latitude is consistent with differential rotation. The dashed vertical lines in each image indicates an integer equatorial rotation.}
    \label{fig.NS_Acorr_sat_comp}
\end{figure}

With the solar equatorial Rossby modes removed, we reconstruct the flow maps by performing the inverse Fourier transform. The maps are then remapped onto the Stoneyhurst coordinate system. Figure~\ref{fig.NS_Acorr_sat_comp} shows the NS auto-correlation maps after removal of the Rossby waves.

While this filtering works near the equator, at higher latitudes the signal of the long-lived flows also becomes filtered. This is because the Rossby wave frequencies are independent of the differential rotation (latitude), while the advected flows are not. At higher $m$ the flow frequencies cross over into the Rossby mode filters at mid- to high-latitudes. This explains the lack of deflection in Fig.~\ref{fig.synRossby}. 
The vorticity of large-scale flows changes sign at the equator. The inflow vorticity is weaker then the divergence \citep[factor 5,][]{hindman_etal_2009} and comparable to the Rossby waves. 
This means with or without filtering, obtaining a mean image of the inflow vorticity through our methodology (in north or southern hemisphere) is difficult.

{}

\section{Mode fitting methodology}\label{sec.modeFits}
The estimates on the parameters $\pmb{\Lambda}_m$ in Eq.~\ref{eq.lorenzFit} are determined by minimizing the negative of the log-likelihood function,
\begin{equation}
    J_{\lambda,m}(\pmb{\Lambda}_m) = \sum\limits^K_{k=1}\ln \mathcal{L}_\lambda(\omega_k,\pmb{\Lambda}_m) + P(\lambda,\omega_k)/\mathcal{L}_\lambda(\omega_k,\pmb{\Lambda}_m),
\end{equation}{}
where $\omega_k$ is the $k$-th frequency bin within the frequency window from $(m-1/2)\Omega_\lambda$ to $(m+1/2)\Omega_\lambda$ consisting of $K$ frequencies. 

The minimization of $J_{\lambda,m}$ and the estimates on the error of the fits are computed from samples drawn by Markov chain Monte Carlo (MCMC) methods.  The priors on the frequencies are $\omega_m \in [m\Omega_\lambda-\Omega_\lambda/2,m\Omega_\lambda+\Omega_\lambda/2]$, where $\Omega_\lambda$ is specified by the Snodgrass rotation rate \citep{snodgrass_ulrich_1990}. The priors on $A_m$ and $B_m$ are such that the MCMC samples are drawn from the set $[0,5\max P(\lambda,\omega)]$. The prior on the line width are $\Gamma_m\in[2(\omega_{k+1}-\omega_k),\Omega_\lambda/2]$. We utilize the python library {emcee} \citep{emcee} with 100 walkers and 1000 steps. The first 200 steps are taken as burn in and we compute the 16\textsuperscript{th}, 50\textsuperscript{th} and 84\textsuperscript{th} percentiles from the remaining 800 steps. {We verified these results with the semi-analytical method of \citet{toutain_appourchaux_1994}. The results obtained through MCMC are presented here since the error calculations are more numerically stable for fits at higher $m$ and $\lambda$.}

\begin{acknowledgements}
\\
CSH thanks Jishnu Bhattacharya for insightful discussions.\\
\textbf{Funding:} The Center for Space Science at  NYU Abu Dhabi is funded by NYUAD Institute Grant G1502. This work was designed at a workshop supported by the TIFR-Max Planck partner group program. LG acknowledges partial support from ERC Synergy Grant WHOLE~SUN 810218.\\
\textbf{Data:} This work utilizes data obtained by the Global Oscillation Network
Group (GONG) program, managed by the National Solar Observatory, which
is operated by AURA, Inc. under a cooperative agreement with the National
Science Foundation. The data were acquired by instruments operated
by the Big Bear Solar Observatory, High Altitude Observatory, Learmonth
Solar Observatory, Udaipur Solar Observatory, Instituto de Astrof{\`i}sica
de Canarias, and Cerro Tololo Interamerican Observatory. The HMI data is courtesy
of NASA/SDO and the HMI Science Team. The HMI LCT maps are courtesy of Bj{\"o}rn L{\"o}ptien. The results presented here may be obtained from the authors upon request.\\
\textbf{Software:} In this study we used python 3.6.7\footnote{www.python.org/downloads/release/python-367/}, with the packages numpy 1.17.2\footnote{numpy.org/devdocs/release/1.17.2-notes.html}, scipy 1.3.1\footnote{docs.scipy.org/doc/scipy/reference/release.1.3.1.html} and astropy 3.2.2\footnote{www.astropy.org/}. Processing of MDI and HMI data to produce data that is not available from the JSOC pipeline was performed using NetDRMS on the compute cluster at MPS and NYUAD.
\end{acknowledgements}

 \bibliographystyle{aa}
\bibliography{References}

\end{document}